\documentclass[12pt]{article}

\usepackage{graphicx}
\usepackage{amsmath}
\usepackage{amssymb}
\allowdisplaybreaks

\begin{document}

\baselineskip=17.5pt plus 0.2pt minus 0.1pt

\renewcommand{\theequation}{\arabic{equation}}
\renewcommand{\thefootnote}{\fnsymbol{footnote}}
\makeatletter
\def\CR{\nonumber \\}
\def\be{\begin{equation}}
\def\ee{\end{equation}}
\def\bea{\begin{eqnarray}}
\def\eea{\end{eqnarray}}
\def\bead{\be\begin{aligned}}
\def\eead{\end{aligned}\ee}
\def\eq#1{(\ref{#1})}
\def\la{\langle}
\def\ra{\rangle}
\def\hyp{\hbox{-}}
\def\ul{\underline}

\begin{titlepage}
\title{\hfill\parbox{4cm}{ \normalsize YITP-11-62}\\
\vspace{1cm} Super tensor models, super fuzzy spaces \\
and super n-ary transformations}
\author{
Naoki {\sc Sasakura}\thanks{\tt sasakura@yukawa.kyoto-u.ac.jp}
\\[15pt]
{\it Yukawa Institute for Theoretical Physics, Kyoto University,}\\
{\it Kyoto 606-8502, Japan}}
\date{}
\maketitle
\thispagestyle{empty}
\begin{abstract}
\normalsize
By extending the algebraic description of 
the bosonic rank-three tensor models,  
a general framework for super rank-three tensor models and 
correspondence to super fuzzy spaces is proposed.
The corresponding super fuzzy spaces must satisfy a certain cyclicity condition on the 
algebras of functions on them.
Due to the cyclicity condition, the symmetry of the super rank-three tensor models are 
represented by super n-ary transformations.
The Leibnitz rules and the fundamental identities for the super n-ary transformations
are discussed from the perspective of the symmetry of 
the algebra of a fuzzy space.
It is shown that the 
super n-ary transformations of finite orders which conserve the algebra of a fuzzy space
form a finite closed n-ary super Lie algebra. 
Super rank-three tensor models would be of physical interest as background independent models 
for dynamical generation of supersymmetric fuzzy spaces, in which quantum corrections are under control.
\end{abstract}
\end{titlepage}

\section{Introduction}
\label{sec:intro}

Tensor models have originally been introduced \cite{Ambjorn:1990ge,Sasakura:1990fs,Godfrey:1990dt}
to describe the simplicial quantum gravity in general dimensions higher than two,
with the hope to extend the success of the matrix models in the study of the two-dimensional
simplicial quantum gravity.
Tensor models have later been extended to describe the spin foam and loop quantum gravities by
considering Lie-group valued indices \cite{Boulatov:1992vp,Ooguri:1992eb,DePietri:1999bx}.
These models with group indices, so called group field theory \cite{Freidel:2005qe,Oriti:2009wn}, 
are actively studied with various interesting recent progress \cite{Bonzom:2011zz}-\cite{Gurau:2009tw}.
While the formal arguments so far have shown the promising features of the tensor models as quantum
gravity, the real connection between the tensor models and gravity has not yet been established, because
of the hard and yet unknown non-perturbative dynamics of the tensor models.
From this perspective,  
it would especially be remarkable that the leading critical behavior of the tensor models has recently 
been analyzed \cite{Bonzom:2011zz,Carrozza:2011jn,Gurau:2011xq,Gurau:2010ba} 
in the framework of the colored tensor models \cite{Gurau:2009tw}.
 
In this paper, I will consider super extension of the tensor models. 
One of the motivations to consider such super extension is to circumvent the 
above complications of the non-perturbative dynamics of the tensor models.
In supersymmetric field theories, there exist various examples in which  
quantum corrections of bosonic and fermionic degrees of freedom cancel with each other so that
the full dynamics of the theory is dramatically simplified enough to be studied analytically. 
Similarly, it would be possible that one can construct supersymmetric tensor models
which can well be analyzed semi-classically as in \cite{Sasakura:2006pq}-\cite{Sasakura:2010rb}.  
In the semi-classical analysis of these papers, it has been argued that
the low-lying low-momentum modes around the classical
solutions corresponding to the flat spaces 
in the rank-three tensor models agree exactly with 
the modes of the general relativity or the scalar-tensor theory of gravity.
Therefore supersymmetric tensor models may provide concrete examples in which
the connection between the rank-three tensor models and gravity 
can explicitly be given. 

Another motivation is to consider matters in the context of the tensor models.
An interesting possibility in quantum gravity would be obtaining fermions
as spinorial configurations of bosonic degrees of freedom \cite{Friedman:1982du}.  
However, in view of the present status of the rather insufficient understanding of the dynamics of the tensor models, 
arguments in this direction would become too speculative. 
A more tractable direction would be to introduce 
Grassmann odd variables to the tensor models. 
As seen in the sequel, this leads naturally to 
supersymmetric extension of the tensor models.
By considering appropriate background solutions of the tensor models, 
one would obtain supersymmetric field theories on supersymmetric fuzzy spaces. 
This way of constructing ``our world" would also be in accord with the phenomenological
requirement of supersymmetry in the GUT scenarios \cite{Nakamura:2010zzi}.     

In fact, there exist multiple ways how to introduce Grassmann odd variables to the tensor models. 
In this paper, I will only consider the rank-three tensor models which have a three-index tensor
as their only dynamical variable, and use the correspondence between the rank-three tensor models and
fuzzy spaces, which has been developed in \cite{Sasakura:2011ma,Sasakura:2005js}.
The extension of bosonic fuzzy spaces to super fuzzy spaces will naturally lead to super tensor models.
An advantage of this approach is that the super tensor models so constructed naturally incorporate
supersymmetries. Therefore it would be possible that the classical solutions of 
the super rank-three tensor models will generate various supersymmetric fuzzy spaces which have so far
been constructed based on super Lie algebras \cite{Grosse:1995pr}-\cite{DeBellis:2010pf}. 

In the previous papers \cite{Sasakura:2011ma,Sasakura:2011nj}, it has been shown that
n-ary transformations generate the symmetry transformations of the bosonic rank-three tensor models.
The study with n-ary algebras in physics has been initiated by Nambu \cite{Nambu:1973qe},
who considered a 3-ary extension of the Poisson bracket to generalize Hamilton dynamics.
The fundamental properties of the n-ary Nambu bracket have been formulated in
\cite{Takhtajan:1993vr}, and the super extension has been considered in \cite{Sakakibara}.
The n-ary algebras have recently attracted much attention in the context of M-theory 
\cite{Bagger:2006sk,Gustavsson:2007vu,Bagger:2007jr}, and 
have also been derived in describing some symmetric fuzzy spaces \cite{DeBellis:2010pf}.
Thus the n-ary algebras provide new approaches in the study of quantum spaces, and  
in this paper, the previous bosonic results on n-ary transformations in \cite{Sasakura:2011ma,Sasakura:2011nj}
will be extended to the super case.

This paper is organized as follows.
In the following section, a general framework for 
super fuzzy spaces and correspondence to the rank-three tensor models is presented.
A cyclicity condition on the algebra of functions on a super fuzzy space corresponds to 
the generalized hermiticity condition in the super tensor models.
In Section \ref{sec:transbasis}, super transformations of basis functions are considered, and transformation
properties of tensors are derived for later use.
In Section \ref{sec:metriccons}, it is shown that,
due to the cyclicity condition on the algebra of a super fuzzy space, 
there exists a systematic construction of the n-ary transformations which conserve the metric
of the super algebra.
These transformations correspond to the symmetry transformations of the super rank-three tensor models.
In Section \ref{sec:invariance}, Leibnitz rules and fundamental identities for the n-ary transformations 
are discussed from the perspective of the symmetry of the algebra of a super fuzzy space. 
It is shown that the n-ary transformations of finite orders which conserve the algebra
of a super fuzzy space will form a finite closed n-ary super Lie algebra. 
In Section \ref{sec:action}, a general method of constructing actions of the rank-three tensor
models is discussed.
The final section is devoted to summary and future prospects.

\section{Super fuzzy spaces and super tensor models}
The fundamental ingredients of a super fuzzy space are supernumbers \cite{DeWitt:1992cy} and superfunctions. 
They are separated into two classes, even and odd supernumbers and superfunctions. For a supernumber $\xi$ and 
a superfunction $\phi_a$, one may assign $Z_2$ grades by 
\bea
\underline{\xi},\underline{\phi_a}=
\left\{
\begin{array}{ll}
0 & \hbox{for even supernumber and superfunction},\\
1 & \hbox{for odd supernumber and superfunction}.
\end{array}
\right.
\eea
The $Z_2$ grade for a superfunction will be abbreviated by using its index as
\be
\underline{a}=\underline{\phi_a}
\ee
in this paper. 
With the $Z_2$ grade, 
the commutation between two supernumbers and between a supernumber and a superfunction 
is given by
\bead
\label{eq:comxi}
\xi_1 \xi_2&=(-1)^{\underline{\xi_1}\,\underline{\xi_2}}\xi_2\xi_1, \\
\xi \phi_a &= (-1)^{\underline{a}\,\underline{\xi}}\phi_a \xi.
\eead
Commutation between two superfunctions does not have such a simple relation in general.  

Let me introduce the complex conjugation\footnote{In the previous papers \cite{Sasakura:2011ma,Sasakura:2011nj},
the operation of the complex conjugation was not considered explicitly. This is because both the structure
constants and the symmetry transformations can be consistently taken to be real. In the super case, however, this
is not possible, because super transformations contain imaginary numbers, as can be seen in the sequel. The 
framework in this paper can be applied to the pure bosonic case, and gives another but physically equivalent
correspondence of the rank-three tensor models to fuzzy spaces.},
which satisfies  
\bead
\label{eq:cpconj}
(\xi_1 \xi_2)^*&=\xi_2^* \xi_1^*, \\
(\xi \phi_a)^*&= \phi_a^* \xi^*, \\
(\phi_a \phi_b)^*&=\phi_b^* \phi_a^*.
\eead
The $Z_2$ grade is assumed to be unchanged under the complex conjugation.
In this paper, real super fuzzy spaces are considered: the basis functions $\{ \phi_a|a=1,2,\ldots,N\}$
on a fuzzy space are assumed to satisfy
\be
\phi_a^*=\phi_a.
\label{eq:realbasis}
\ee

The structure of a fuzzy space is characterized by the algebra of the basis functions, 
\be
\phi_a \phi_b =f_{ab}{}^c \phi_c.
\label{eq:fuzzyalg}
\ee
The associativity of the algebra is not assumed in general. The structure constant $f_{ab}{}^c$ 
takes supernumbers in general, 
and the consistency of \eq{eq:fuzzyalg} with the $Z_2$ grade requires 
\be
\underline{f_{ab}{}^c}=\underline{a}+\underline{b}+\underline{c}\  \ \hbox{mod 2}.
\label{eq:z2f}
\ee
In addition, the consistency between \eq{eq:realbasis} and \eq{eq:fuzzyalg} requires 
\be
{f_{ab}{}^c}^*=(-1)^{\ul{c}\,(\ul{a}+\ul{b}+\ul{c})} f_{ba}{}^c.
\label{eq:fstar}
\ee

The algebra is assumed to have a metric, 
\be
\langle \phi_a | \phi_b \rangle =h_{ab},
\label{eq:defh}
\ee
where $h_{ab}$ take supernumbers, and the grade is given by
\be
\ul{h_{ab}}=\ul{a}+\ul{b} \ \ \hbox{mod 2}.
\label{eq:gradeh}
\ee 
The metric \eq{eq:defh} is assumed to be bilinear\footnote{
In this paper, the complex conjugation is not applied to the bra state unlike what is usual in quantum mechanics.
This convention is simple and natural, because the functions are real in the present case.
Otherwise, the complex conjugation would complicate the rules of signatures. See for example \cite{Scheunert:1976wi}.} 
with signatures
inheriting from \eq{eq:comxi}, such as
\bead
\langle \xi_1 \phi_a +\xi_2 \phi_b | \phi_c \rangle&=\xi_1 \la \phi_a |\phi_c \ra + \xi_2 \la \phi_b | \phi_c \ra, \\
\langle \phi_a | \xi_1 \phi_b +\xi_2 \phi_c \rangle&=(-1)^{\ul{\xi_1}\, \ul{a}}
\xi_1 \la \phi_a |\phi_b \ra + (-1)^{\ul{\xi_2}\,\ul{a}} \xi_2 \la \phi_a | \phi_c \ra.
\label{eq:bilinear}
\eead
It is also assumed that the metric is consistent with the grade and the complex conjugation as 
\bead
h_{ab}^*&=\langle \phi_a | \phi_b \rangle^*=\langle \phi_b^* | \phi_a^* \rangle=h_{ba} \\
h_{ab}&=\langle \phi_a | \phi_b \rangle=(-1)^{\ul{a}\,\ul{b}} \langle \phi_b | \phi_a \rangle=(-1)^{\ul{a}\,\ul{b}} h_{ba}
\label{eq:hco}
\eead
In particular, these conditions imply that the even-even elements of $h_{ab}$ take real values, 
while the odd-odd elements take pure imaginary values.

The cyclicity condition on the algebra discussed previously in the bosonic case 
\cite{Sasakura:2011ma,Sasakura:2011nj} can be extended to
\bead
\langle \phi_a \phi_b|\phi_c\rangle=\langle \phi_a|\phi_b \phi_c \rangle=(-1)^{\ul{c}(\ul{a}+\ul{b})}\langle 
\phi_c \phi_a | \phi_b
\rangle = (-1)^{\ul{c}(\ul{a}+\ul{b})} \langle \phi_c | \phi_a \phi_b \rangle  \\
=(-1)^{\ul{a}(\ul{b}+\ul{c})}\langle \phi_b \phi_c | \phi_a \rangle
=(-1)^{\ul{a}(\ul{b}+\ul{c})} \langle \phi_b | \phi_c \phi_a \rangle.
\label{eq:cyclicalg}
\eead 
The tensor models considered in this paper are 
the rank-three tensor models which have a three-index tensor $M_{abc}$ 
as their only dynamical variable.  
As in \cite{Sasakura:2011ma,Sasakura:2011nj}, the correspondence between the tensor models and 
the fuzzy spaces is assumed to be
\be
M_{abc}=f_{ab}{}^d h_{dc}.
\label{eq:Mfrel}
\ee 
Then one obtains
\be
M_{abc}=(-1)^{\ul{a}(\ul{b}+\ul{c})}M_{bca}=(-1)^{\ul{c}(\ul{a}+\ul{b})}M_{cab}
=M_{cba}^*=(-1)^{\ul{a}(\ul{b}+\ul{c})}M_{acb}^*=(-1)^{\ul{c}(\ul{a}+\ul{b})}M_{bac}^*.
\label{eq:genher}
\ee
from the properties of the algebra.
This is the super extension of the generalized hermiticity condition in the tensor
models.

\section{Super transformations of basis functions}
\label{sec:transbasis}
Let me consider the following infinitesimal transformation of basis functions,
\be
\delta \phi_{a}=T_{a}{}^b \phi_b,
\label{eq:phip}
\ee
where $T_{a}{}^b$ take supernumbers. 
The consistency of the grade in \eq{eq:phip} requires
\be
\ul{T_a{}^b}=\ul{a} + \ul{b}\ \ \hbox{mod 2}.
\ee 
The reality condition \eq{eq:realbasis} requires
\be
T_a{}^b=(-1)^{\ul{b}(\ul{a}+\ul{b})}T_a{}^b{}^*.
\ee

The transformation of the metric $h_{ab}$ by \eq{eq:phip} is given by
\bead
\delta h_{ab}&= \langle \delta \phi_a | \phi_b \rangle + \langle \phi_a | \delta \phi_b \rangle \\
&= T_{a}{}^{a'} h_{a'b}+(-1)^{\ul{a}(\ul{b}+\ul{b'})}T_{b}{}^{b'} h_{ab'}.
\label{eq:transh}
\eead
Similarly, from \eq{eq:fuzzyalg}, one obtains 
\be
\delta f_{ab}{}^c=T_{a}{}^{a'} f_{a'b}{}^c+(-1)^{\ul{a}(\ul{b}+\ul{b'})}T_b{}^{b'} f_{ab'}{}^c
-(-1)^{(\ul{c}+\ul{c'})(\ul{a}+\ul{b}+\ul{c'})}T_{c'}{}^c 
f_{ab}{}^{c'}.
\label{eq:transfup}
\ee
Then, from \eq{eq:Mfrel}, \eq{eq:transh} and \eq{eq:transfup}, one obtains
\be
\delta M_{abc}=T_a{}^{a'} M_{a'bc}+(-1)^{\ul{a}(\ul{b}+\ul{b'})}T_b{}^{b'} M_{ab'c}
+(-1)^{(\ul{a}+\ul{b})(\ul{c}+\ul{c'})} T_c{}^{c'} M_{abc'}.
\label{eq:transm}
\ee

To construct actions of the tensor models, it is necessary to consider the inverse of the metric $h_{ab}$.
It is defined by
\be
h^{ac}h_{cb}=h_{ac}h^{cb}=\delta^K_{ab},
\label{eq:hinv}
\ee
where $\delta^K_{ab}$ denotes the Kronecker delta. 
The unique existence of the inverse under a certain condition is proven
in \cite{DeWitt:1992cy}.
The consistency of the grade in \eq{eq:hinv} requires
\be
\ul{h^{ab}}=\ul{a}+\ul{b}\ \ \hbox{mod 2}.
\label{eq:gradehup}
\ee
The $h^{ab}$ can be shown to have the properties,
\bead
h^{ab}{}^*&=h^{ba},\\
h^{ab}&=(-1)^{\ul{a}+\ul{b}+\ul{a}\,\ul{b}} h^{ba}.
\label{eq:hupco}
\eead
The former equation can be obtained by
taking the complex conjugate of \eq{eq:hinv} and using \eq{eq:hco},
while the latter one can be obtained by exchanging the order of $h^{ab}$ and $h_{ab}$ in \eq{eq:hinv} as 
\bead
\delta^K_{ab}&=h^{ac}h_{cb} \\
&= (-1)^{\ul{a}+\ul{c}+\ul{a}\,\ul{c}} h_{bc} h^{ac},
\label{eq:exorderh}
\eead 
where \eq{eq:gradeh}, \eq{eq:hco}, \eq{eq:hinv} and \eq{eq:gradehup} are used,
and comparing it with \eq{eq:hinv}.
Finally the transformation of $h^{ab}$ by \eq{eq:phip} is given by
\bead
\delta h^{ab}&=-h^{aa'}\delta h_{a'b'} h^{b'b} \\ 
&= -(-1)^{\ul{a}+\ul{a}\,\ul{a'}} T_{a'}{}^a h^{a'b} - (-1)^{(\ul{b}+\ul{b'})(\ul{a}+\ul{b'})} T_{b'}{}^b h^{ab'}.
\label{eq:huptrans}
\eead 

\section{Metric conserving super n-ary transformations}
\label{sec:metriccons}
Let me consider the following sum of products of $n$ functions given by 
\be
(\epsilon,s;\phi_a)_n\equiv (-1)^{\ul{a}(\ul{b_{p}}+\ul{b_{p+1}}+\cdots+\ul{b_{n-1}})}\epsilon^{b_1b_2\ldots b_{n-1}}
\phi_{b_1} \phi_{b_2} \cdots \phi_{b_{p-1}} \phi_a \phi_{b_p} \cdots \phi_{b_{n-1}},
\label{eq:narytrans}
\ee
where $b_i$'s are summed over, and $\epsilon^{b_1b_2\ldots b_{n-1}}$ are supernumbers with grades,
\be
\ul{\epsilon^{b_1b_2\ldots b_{n-1}}}=\sum_{i=1}^{n-1} \ul{b_i}\ \ \hbox{    mod 2}.
\ee 
The equation \eq{eq:narytrans} can be regarded as 
a definition of the infinitesimal linear transformation which transforms the basis functions $\phi_a$ by
\eq{eq:narytrans}.
The reason for the introduction of $\epsilon^{b_1b_2\ldots b_{n-1}}$ into \eq{eq:narytrans}
is to cancel the grades of $\phi_{b_i}$ so that \eq{eq:narytrans} becomes consistent 
as an infinitesimal transformation of the basis functions.
The right-hand side of \eq{eq:narytrans} is not well defined, because 
the order of the products must explicitly be shown by parentheses to take care of 
the possible lack of the associativity of the algebra. 
So the symbol $s$ in \eq{eq:narytrans} is supposed to express the order implicitly.  
The sign factor in \eq{eq:narytrans} 
is necessary for the existence of a transpose linear transformation, which will be defined below. 
Here all the $\ul{a}$ dependence of the sign factor is assumed to be exhausted by this expression.
The other possible sign factors dependent only on $\underline{b_i}$'s  are supposed to be included in
$\epsilon^{b_1b_2\ldots b_{n-1}}$.   

A linear transformation $\overline{(\epsilon,s;\phi_a)_n}$  transpose to $(\epsilon,s;\phi_a)_n$ is defined by
\be
\left\langle \left.\overline{( \epsilon,s;\phi_a)_n}\right| \phi_b \right\rangle=\langle \phi_a | 
(\epsilon,s;\phi_b)_n\rangle.
\label{eq:defofbar}
\ee
The transpose linear transformation can be obtained by transferring $\phi_{b_i}$'s in the ket of the right-hand side of
\eq{eq:defofbar} to the bra by using the cyclicity condition \eq{eq:cyclicalg} of the algebra. In the transfer,
some sign factors are created, but the sign factor in \eq{eq:narytrans} cancel exactly the sign factors 
unfavorable for the transpose to exist. This can be seen in an example below.

Since the bra and ket in \eq{eq:defofbar} can be exchanged due to \eq{eq:hco},
transpososition of transposition is identity,
\be
\overline{\overline {(\epsilon,s;\phi_a)_n}}=(\epsilon,s;\phi_a)_n,
\label{eq:barbar}
\ee

Let me next define 
\be
(\epsilon,s;\phi_a)_{n-}\equiv (\epsilon,s;\phi_a)_n-\overline{(\epsilon,s;\phi_a)_n}.
\ee
From \eq{eq:barbar}, one can easily show that this is a metric conserving infinitesimal 
transformation,
\be
\langle (\epsilon,s;\phi_a)_{n-} |\phi_b \rangle = -\langle \phi_a | (\epsilon,s;\phi_b)_{n-}\rangle.
\ee
A transformation of basis functions must respect the reality condition \eq{eq:realbasis}.  
Therefore one should rather consider
\bead
\delta_{\epsilon,s} \phi_a&\equiv 
(\epsilon,s;\phi_a)_{n-}+(\epsilon,s;\phi_a)_{n-}^* \\
&= (\epsilon,s;\phi_a)_n-\overline{(\epsilon,s;\phi_a)_n}+
(\epsilon,s;\phi_a)_n^*-{\overline{(\epsilon,s;\phi_a)_n}}^{\,*}.
\label{eq:narydelphi}
\eead
The transposition and the complex conjugation commute with each other,
\be
{\overline{(\epsilon,s;\phi_a)_n}}^{\,*}=\overline{(\epsilon,s;\phi_a)_n^*}.
\ee
This can easily be shown by taking the complex conjugation of \eq{eq:defofbar}
after replacing $(\epsilon,s;\phi_a)_n\rightarrow \overline{(\epsilon,s;\phi_a)_n}$. 
Therefore \eq{eq:narydelphi}
has indeed the metric conserving form,
\be
\delta_{\epsilon,s} \phi_a = (\epsilon,s;\phi_a)_n-\overline{(\epsilon,s;\phi_a)_n}+
(\epsilon,s;\phi_a)_n^*-\overline{(\epsilon,s;\phi_a)_n^*},
\label{eq:narydelphi2}
\ee
Thus \eq{eq:narydelphi} or \eq{eq:narydelphi2} gives the super extension of the metric conserving 
infinitesimal n-ary transformation
discussed previously for the bosonic case in \cite{Sasakura:2011ma,Sasakura:2011nj}.

It would be instructive to see some simple examples. Let me first consider
product of two functions,
\be
(\epsilon,s;\phi_a)_2\equiv \epsilon^b \phi_b \phi_a.
\label{eq:extrans2}
\ee
By using the cyclicity condition \eq{eq:cyclicalg} and \eq{eq:bilinear}, one finds
\bead
\langle \phi_a | (\epsilon,s;\phi_c)_2\rangle&=\langle \phi_a | \epsilon^b \phi_b \phi_c \rangle \\
&=(-1)^{\ul{a}\,\ul{b}} \langle \epsilon^b \phi_a \phi_b | \phi_c \rangle.
\eead
Therefore the transpose is given by
\be
\overline{(\epsilon,s;\phi_a)_2}=(-1)^{\ul{a}\,\ul{b}} \epsilon^b \phi_a \phi_b.
\ee
Then, from \eq{eq:cpconj}, one obtains
\bead
(\epsilon,s;\phi_a)_2^*&= (-1)^{\ul{b}+\ul{a}\,\ul{b}}\, \epsilon^b{}^*\phi_a \phi_b, \\
\overline{(\epsilon,s;\phi_a)_2}^{\,*}&= (-1)^{\ul{b}}\, \epsilon^b{}^* \phi_b \phi_a.
\eead
Thus, from \eq{eq:narydelphi}, the metric conserving infinitesimal linear transformation of the
basis functions is obtained as  
\be
\delta_{\epsilon,s}\phi_a
=\left(\epsilon^b-(-1)^{\ul{b}} \epsilon^b{}^*\right) \{ [\phi_b,\phi_a]\},
\ee
where $\{ [\ ,\ ]\}$ is the supercommutator,
\be
\{ [\phi_a,\phi_b]\} \equiv \phi_a \phi_b-(-1)^{\ul{a}\,\ul{b}} \phi_b\phi_a.
\ee

As another example, let me consider the following product of three functions,
\be
(\epsilon,s;\phi_a)_3\equiv (-1)^{\ul{a}\,\ul{b_2}}\epsilon^{b_1b_2}
(\phi_{b_1}\phi_a) \phi_{b_2},
\label{eq:s3def}
\ee
where the sign factor follows \eq{eq:narytrans}. Here the order of product is 
explicitly indicated by the parentheses. Transfer of $\phi_{b_i}$ 
by using the cyclicity condition \eq{eq:cyclicalg}
leads to
\bead
\la \phi_a | (\epsilon,s;\phi_c)_3\ra&= 
\la \phi_a | (-1)^{\ul{c}\,\ul{b_2}}\epsilon^{b_1b_2}
(\phi_{b_1}\phi_c) \phi_{b_2} \rangle \\
&=  (-1)^{\ul{c}\,\ul{b_2}+(\ul{b_1}+\ul{b_2})\ul{a} + 
\ul{b_2} (\ul{a}+\ul{b_1}+\ul{c})} \epsilon^{b_1 b_2} \la (\phi_{b_2} \phi_a) \phi_{b_1} | \phi_c \ra.
\eead
Note that, in the last line, the dependence on $\underline{c}$ of the sign factor cancels out. Therefore one can
consistently obtain the transpose linear transformation as
\be
\overline{(\epsilon,s_3;\phi_a)_3}=(-1)^{\ul{a}\,\ul{b_1} +\ul{b_1}\,\ul{b_2}}\,
\epsilon^{b_1b_2} (\phi_{b_2} \phi_a)\phi_{b_1}.
\ee
If the sign factor in \eq{eq:s3def} has not been in this form, the dependence on $\underline{c}$ would have remained and
the transpose linear transformation could not have been obtained consistently. 
From this simple example, it is clear that the sign factor needed for the general case is given by the form
in \eq{eq:narytrans}.

Another comment on the sign factor is about the linearity of the transformation \eq{eq:narytrans}.
The $\ul{a}$ in the sign factor of \eq{eq:narytrans} should denote the grade of the whole
expression of the last entry of the left-hand side of \eq{eq:narytrans}. 
With this convention, for example, when the last entry is multiplied by 
a supernumber $\xi$, the following simple linear property holds,
\bead
(\epsilon,s;\xi \phi_a)_n&= (-1)^{(\ul{a}+\ul{\xi})
(\ul{b_{p}}+\ul{b_{p+1}}+\cdots+\ul{b_{n-1}})}\epsilon^{b_1b_2\ldots b_{n-1}}
\phi_{b_1} \phi_{b_2} \cdots \phi_{b_{p-1}} \xi \phi_a \phi_{b_p} \cdots \phi_{b_{n-1}}\\
&= (-1)^{\ul{a}
(\ul{b_{p}}+\ul{b_{p+1}}+\cdots+\ul{b_{n-1}})} \xi \epsilon^{b_1b_2\ldots b_{n-1}}
\phi_{b_1} \phi_{b_2} \cdots \phi_{b_{p-1}} \phi_a \phi_{b_p} \cdots \phi_{b_{n-1}}\\
&=\xi\, (\epsilon,s;\phi_a)_n. 
\label{eq:linnaryexam}
\eead
This convention for the sign factor should also be applied when the last entry is a product of functions $\phi_a$.

In relation to the above linearity, it is more appropriate to rewrite \eq{eq:narydelphi}
or \eq{eq:narydelphi2} in the form,   
\be
\delta_{\epsilon,s} \phi_a= (\epsilon,s;\phi_a)_n-(\overline{\epsilon},\overline{s};\phi_a)_n+
(\epsilon^*,s^*;\phi_a)_n-(\overline{\epsilon}^*,\overline{s}^*;\phi_a)_n,
\label{eq:lineardelta}
\ee
where $\overline{\epsilon},\overline{s},\epsilon^*,s^*$ are defined by 
\bead
(\overline{\epsilon},\overline{s};\phi_a)_n&=\overline{(\epsilon,s;\phi_a)}_n,\\
(\epsilon^*,s^*;\phi_a)_n&=(\epsilon,s;\phi_a)_n^*.
\label{eq:sbarsstar}
\eead
The consistency of the definition \eq{eq:sbarsstar} comes from the fact that $\overline{(\epsilon,s;\phi_a)_n}$ and 
$(\epsilon,s;\phi_a)_n^*$ have also the form of \eq{eq:narytrans} with appropriate replacement of $\epsilon$ and $s$.
The $\delta_{\epsilon,s} \phi_a$ in the form \eq{eq:lineardelta} respects the linearity,
\be
\delta_{\epsilon,s}(\xi \phi_a)=\xi\, \delta_{\epsilon,s} \phi_a,
\ee
while the form \eq{eq:narydelphi} or \eq{eq:narydelphi2} does not. 

\section{Invariance of algebra, Leibnitz rule, and closed n-ary Lie algebra}
\label{sec:invariance}

Let me consider an infinitesimal linear transformation which is given by 
a finite sum of n-ary transformations,
\bead
\delta_L \phi_a&=\sum_{n\leq n_{max}}\sum_{\epsilon,s} (\epsilon,s;\phi_a)_n \\
&= L_{a}{}^{a'} \phi_{a'},
\label{eq:sumnary}
\eead
where $L_a{}^{a'}$ are supernumbers. From \eq{eq:sumnary}, the grade of $L_a{}^{a'}$ is given by
\be
\ul{L_{a}{}^{a'}}=\underline{a}+\ul{a'}.
\ee 
The form in \eq{eq:sumnary} contains the metric conserving real transformations \eq{eq:lineardelta}
as special cases, but more general cases are allowed in the following discussions. 

Now let me assume that $L_{a}{}^{a'}$ satisfies
\be
L_{a}{}^{a'} f_{a'b}{}^c+(-1)^{\ul{a}(\ul{b}+\ul{b'})}L_b{}^{b'} f_{ab'}{}^c
-(-1)^{(\ul{c}+\ul{c'})(\ul{a}+\ul{b}+\ul{c'})}L_{c'}{}^c 
f_{ab}{}^{c'}=0,
\label{eq:consalg}
\ee
which is nothing but the condition that the algebra \eq{eq:fuzzyalg}  is invariant under the
infinitesimal transformation \eq{eq:sumnary} (See \eq{eq:transfup}).
Then it is straightforward to show that the transformation \eq{eq:sumnary} satisfies the Leibnitz rule,
\be
\delta_L(\phi_a \phi_b) =\delta_L(\phi_a) \phi_b+\phi_a \delta_L(\phi_b).
\label{eq:leibnitz}
\ee
In the proof, it is essentially important to take the convention of the sign factor taken for example
in \eq{eq:linnaryexam} to guarantee the linearity. 
By iteratively applying \eq{eq:leibnitz}, it is obvious that the Leibnitz
rule holds for any product of superfunctions.

Let me apply $\delta_L$ to an arbitrary n-ary transformation \eq{eq:narytrans}.
Then one obtains the so called fundamental identity,
\bead
\label{eq:comnarydel}
\delta_L\, (\epsilon,s,\phi_a)_n&=\delta_L \left( (-1)^{\ul{a}(\ul{b_{p}}+\ul{b_{p+1}}+\cdots+\ul{b_{n-1}})}
\epsilon^{b_1b_2\ldots b_{n-1}}
\phi_{b_1} \phi_{b_2} \cdots \phi_{b_{p-1}} \phi_a \phi_{b_p} \cdots \phi_{b_{n-1}}\right) \\
&=(-1)^{\ul{a}(\ul{b_{p}}+\cdots+\ul{b_{n-1}})}
\epsilon^{b_1\ldots b_{n-1}}
\left(
\delta_L(\phi_{b_1}) \phi_{b_2} \cdots \phi_{b_{p-1}} \phi_a \phi_{b_p} \cdots \phi_{b_{n-1}} \right. \\
&\left.\ \ \ \ \ \ \ +\phi_{b_1} \delta_L(\phi_{b_2}) \cdots \phi_{b_{p-1}} \phi_a \phi_{b_p} \cdots \phi_{b_{n-1}}
+\cdots \right)\\
&=(-1)^{\ul{a}(\ul{b_{p}}+\cdots+\ul{b_{n-1}})}
\epsilon^{b_1 \ldots b_{n-1}}
\left(
L_{b_1}{}^{b_1'} \phi_{b_1'} \phi_{b_2} \cdots \phi_{b_{p-1}} \phi_a \phi_{b_p} \cdots \phi_{b_{n-1}} \right. \\
&\ \ \ \ \ \ \ +\phi_{b_1} L_{b_2}{}^{b_2'}\phi_{b_2'} 
\cdots \phi_a \cdots \phi_{b_{n-1}}+\cdots 
+\phi_{b_1} \cdots \delta_L(\phi_a)  \cdots \phi_{b_{n-1}} \\
&\left.\ \ \ \ \ \ \ +\phi_{b_1} \cdots \phi_a L_{b_p}{}^{b_p'} \phi_{b_p'} 
 \cdots \phi_{b_{n-1}}+\cdots \right)\\
&=(-1)^{\ul{a}(\ul{b_{p}}+\cdots+\ul{b_{n-1}})}
\epsilon^{b_1 \ldots b_{n-1}}
\left(
L_{b_1}{}^{b_1'} \phi_{b_1'} \phi_{b_2} \cdots \phi_{b_{p-1}} \phi_a \phi_{b_p} \cdots \phi_{b_{n-1}} \right. \\
&\ \ \ \ \ \ \ +(-1)^{\ul{b_1}(\ul{b_2}+\ul{b_2'})}L_{b_2}{}^{b_2'}\phi_{b_1} \phi_{b_2'} 
\cdots \phi_a \cdots \phi_{b_{n-1}}+\cdots 
+\phi_{b_1} \cdots \delta_L(\phi_a)  \cdots \phi_{b_{n-1}} \\
&\left.\ \ \ \ \ \ \ +(-1)^{(\ul{b_1}+\cdots+\ul{b_{p-1}}+\ul{a})(\ul{b_p}+\ul{b_p'})}
L_{b_p}{}^{b_p'} \phi_{b_1} \cdots \phi_a \phi_{b_p'} 
 \cdots \phi_{b_{n-1}}+\cdots \right)\\ 
&=(\epsilon,s;\delta_L(\phi_a))_n+\sum_{i=1}^{n-1} (\epsilon_{L\,i},s;\phi_a)_n,
\eead
where
\be
\epsilon_{L\,i}^{b_1 b_2 \ldots b_{n-1}}=(-1)^{(\ul{b_1}+\ul{b_2}+\cdots+\ul{b_{i-1}})(\ul{b_i}+\ul{b_i'})}
\epsilon^{b_1b_2\ldots b_i' \ldots b_{n-1}} L_{b_i'}{}^{b_i}.
\label{eq:eprime}
\ee

Let me consider $\delta_{L_1}$ and $\delta_{L_2}$ which have the form \eq{eq:sumnary} and satisfy the algebra
conserving condition \eq{eq:consalg}. 
Since it can be shown that the infinitesimal transformations satisfying \eq{eq:consalg} form a Lie algebra,
the commutator $[\delta_{L_1},\delta_{L_2}]$ conserves the algebra. 
In addition, from \eq{eq:comnarydel}, one obtains
\bead
\delta_{[L_1,L_2]} \phi_a 
&=\delta_{L_1}\left( \delta_{L_2} \phi_a \right)-\delta_{L_2}\left( \delta_{L_1} \phi_a \right)\\
&=\sum_{n\leq n_{max}} \sum_{\epsilon',s} (\epsilon',s;\phi_a)_n, 
\label{eq:coml1l2}
\eead
where $\epsilon'$ are computed in the same manner as in \eq{eq:eprime}.
Thus, the infinitesimal 
transformations which have the form \eq{eq:sumnary} and conserve the algebra of a super fuzzy space
will form a closed finite n-ary super Lie algebra.

\section{Actions of super tensor models}
\label{sec:action}

The correspondence between the tensor models and the fuzzy spaces is assumed to be given by \eq{eq:Mfrel}.
While the tensor models have the dynamical degrees of freedom $M_{abc}$, the fuzzy spaces have more
degrees of freedom $f_{ab}{}^c$ and $h_{ab}$.
To balance these degrees of freedom, $h_{ab}$ 
are gauge fixed to constant values in \cite{Sasakura:2011ma,Sasakura:2011nj}
by using the transformations of the basis functions. 
Assuming the same procedure for the super tensor models, 
an action of a super rank-three tensor model is given by a quantity which depends on $h^{ab}$ and $M_{abc}$ 
and is invariant under \eq{eq:transm} and \eq{eq:huptrans}.
Here $h^{ab}$ should be assumed to be non-dynamical, and hence the symmetry of the super tensor models is
given by the metric conserving super n-ary transformations discussed in Section \ref{sec:metriccons}.

Because of the sign factors, the construction of such an action invariant under \eq{eq:transm} and \eq{eq:huptrans} 
directly from $h_{ab}$ and $M_{abc}$ becomes rather complicated in general. 
This becomes much easier by using the algebraic construction discussed in \cite{Sasakura:2011ma}.  
Let me first consider
\be
\phi_a h^{ab} \phi_b.
\label{eq:phihphi}
\ee
This is invariant under \eq{eq:phip}, because
\bead
\delta(\phi_a h^{ab} \phi_b) 
&=(\delta \phi_a) h^{ab} \phi_b+\phi_a (\delta h^{ab}) \phi_b+ \phi_a h^{ab} (\delta \phi_b)\\
&=T_a{}^{a'} \phi_{a'} h^{ab}\phi_b+\phi_a 
\left( -(-1)^{\ul{a}+\ul{a}\,\ul{a'}} T_{a'}{}^a h^{a'b} - (-1)^{(\ul{b}+\ul{b'})(\ul{a}+\ul{b'})} T_{b'}{}^b h^{ab'} 
\right) \phi_b
+\phi_a h^{ab} T_b{}^{b'} \phi_{b'}\\
&=0,
\eead
where \eq{eq:huptrans} is used.
From  
\be
\xi_1 \phi_a h^{ab} \phi_b \xi_2=(-1)^{\ul{a}\,\ul{\xi_1}+\ul{b}\,\ul{\xi_2}}\phi_a \xi_1 h^{ab} \xi_2 \phi_b, 
\label{eq:phixi}
\ee
where $\xi_i$'s are supernumbers, it is clear that
the invariance will be kept by a sign factor which is needed to 
rearrange a given order of an expression to the canonical order of \eq{eq:phihphi},
when there are some insertions of supernumbers. In other words, the sign factor 
cancels exactly the unfavorable sign factors which are created when 
$T_a{}^b$ are reordered in the proof of invariance under \eq{eq:transm} and \eq{eq:huptrans}.
Since $T_a{}^b$ are supernumbers, the insertions can be some functions $\phi_a$'s, provided that
their indices are properly contracted by $h^{ab}$. 
In this way, one can easily construct various invariants, such as   
\bead
&\la \phi_a h^{ab} | \phi_b \ra, \\
&(-1)^{\ul{b}\,\ul{c}}\la \phi_a h^{ab} \phi_c |  \phi_b h^{cd} \phi_d \ra, \\
&(-1)^{\ul{a}\,\ul{b}+\ul{b}\,\ul{c}+\ul{c}\,\ul{a}} 
\la (\phi_a \phi_b) \phi_c | h^{aa'}\phi_{a'}h^{bb'}(\phi_{b'}h^{cc'}\phi_{c'})\ra.
\label{eq:actionalg}
\eead
In fact, the above simple expressions may become quite complicated, if they are written in terms of $M_{abc}$.
For example, the second one can be rewritten as
\bead
(-1)^{\ul{b}\,\ul{c}}\la \phi_a h^{ab} \phi_c |  \phi_b h^{cd} \phi_d \ra 
&=(-1)^{\ul{b}\,\ul{c}+\ul{a}(\ul{a}+\ul{b})+(\ul{a}+\ul{b}+\ul{c})(\ul{c}+\ul{d})}
h^{ab}h^{cd}\la \phi_a \phi_c | \phi_b \phi_d\ra \\
&=(-1)^{\ul{b}\,\ul{c}+\ul{a}(\ul{a}+\ul{b})+(\ul{a}+\ul{b}+\ul{c})(\ul{c}+\ul{d})}
h^{ab}h^{cd} \la f_{ac}{}^e \phi_e | f_{bd}{}^f \phi_f \ra \\
&=(-1)^{\ul{b}\,\ul{c}+\ul{a}(\ul{a}+\ul{b})+(\ul{a}+\ul{b}+\ul{c})(\ul{c}+\ul{d})+
\ul{e}(\ul{b}+\ul{d}+\ul{f})}
h^{ab}h^{cd} f_{ac}{}^e f_{bd}{}^f h_{ef} \\
&= (-1)^{\ul{b}\,\ul{c}+\ul{a}(\ul{a}+\ul{b})+(\ul{a}+\ul{b}+\ul{c})(\ul{c}+\ul{d})+
\ul{e}(\ul{b}+\ul{d})}h^{ab}h^{cd} M_{acf}h^{fe} M_{bde}. 
\eead
Thus the algebraic description of actions like \eq{eq:actionalg} 
is generally much simpler than the expressions directly in terms of the dynamical variable
$M_{abc}$.
This is an advantage of the algebraic description of the super tensor models.

\section{Summary and future prospects}
\label{sec:summary}
In this paper, a general framework for the super rank-three tensor models and 
the correspondence to the super fuzzy spaces has been presented.
As in the bosonic case, the algebras of functions on super fuzzy spaces must satisfy
the cyclicity condition, which corresponds to 
the generalized hermiticity condition of the super tensor models.
This cyclic property enables the systematic construction of the metric conserving 
super n-ary transformations on a super fuzzy space, which are also the symmetry transformations
of the super rank-three tensor models.
In the bosonic case, such metric conserving transformations constitute the 
unitary transformations of a fuzzy space containing the transformations which correspond to
the diffeomorphism on a usual space \cite{Sasakura:2009hs}. 
This would suggest that the super n-ary transformations contain local supersymmetries, which are
the gauge symmetries of supergravities.  
Thus it would be interesting to study the physical roles of the super n-ary transformations by
considering some examples of supersymmetric fuzzy spaces in the framework of this paper.
 
In Section \ref{sec:invariance}, it has been shown that the n-ary transformations of 
finite orders which  conserve the algebra of a fuzzy space will 
form a closed finite super Lie algebra. Such finite n-algebraic relations will provide 
characterizations of some symmetric super fuzzy spaces as in \cite{DeBellis:2010pf}.
It would be interesting to consider some examples in the framework of this paper. 

A method of systematic constructions of the actions of the super rank-three tensor models has been given. 
In the actions, there are sign factors, which are necessary for the invariance under the super transformations.
The sign factors raise the question of stability of the super tensor models. 
But this does not necessarily reduce the physical interests of the models.
In fact, in the matrix models, local minima of potentials are enough to obtain physical 
outcomes in scaling limits. 
One may expect that similar arguments may take care of the situations of the super tensor models,
and may also expect that supersymmetries can stabilize local minima physically enough.
Then the real challenge would be to find local minima corresponding to supersymmetric fuzzy spaces. 
In view of the fact that general relativity is emergent on fuzzy spaces in the bosonic case 
\cite{Sasakura:2007sv}-\cite{Sasakura:2010rb}, it would be highly interesting to
study the emergent field theories on such supersymmetric fuzzy spaces. One may hopefully
find emergent supergravities.


\end{document}